\documentclass[mnsc,nonblindrev]{informs3_homepage}

\OneAndAHalfSpacedXI

\usepackage{xcolor}
\definecolor{pastelBlue}{rgb}{0.0,0.4,0.7}
\usepackage{hyperref}
\hypersetup{
colorlinks=true,
allcolors = pastelBlue 
}

\usepackage{natbib}
 \bibpunct[, ]{(}{)}{,}{a}{}{,}%
 \def\bibfont{\footnotesize}%

\TheoremsNumberedThrough     
\ECRepeatTheorems

\EquationsNumberedThrough

\usepackage{float}
\usepackage[caption=false]{subfig}
\usepackage{amsmath}
\usepackage{algorithm2e}
\RestyleAlgo{ruled}
\LinesNumbered
\let\oldnl\nl
\newcommand{\nonl}{\renewcommand{\nl}{\let\nl\oldnl}}

\usepackage{algpseudocode}
\usepackage{ifthen}
\usepackage{enumitem}
\usepackage{cases}
\usepackage{mathtools}
\usepackage{amssymb}
\usepackage{graphicx}
\usepackage{empheq}
\usepackage{bm}
\usepackage{dsfont}
\usepackage{tikz}
\usetikzlibrary{calc,arrows.meta, positioning, shapes.geometric, shapes.misc, fit, backgrounds}

\usepackage{array,tabularx,booktabs}
\newcolumntype{Y}{>{\raggedright\arraybackslash}X}

\begin{document}

\RUNTITLE{LLM Election}

\TITLE{Simulating and Experimenting with Social Media Mobilization Using LLM Agents}

\ARTICLEAUTHORS{%
\AUTHOR{Sadegh Shirani~~~~~~~and~~~~~~~Mohsen Bayati}
\AFF{Graduate School of Business, Stanford University}
}
\ABSTRACT{%
Online social networks have transformed the ways in which political mobilization messages are disseminated, raising new questions about how peer influence operates at scale. Building on the landmark 61-million-person Facebook experiment \citep{bond201261}, we develop an agent-based simulation framework that integrates real U.S. Census demographic distributions, authentic Twitter network topology, and heterogeneous large language model (LLM) agents to examine the effect of mobilization messages on voter turnout. Each simulated agent is assigned demographic attributes, a personal political stance, and an LLM variant (\texttt{GPT-4.1}, \texttt{GPT-4.1-Mini}, or \texttt{GPT-4.1-Nano}) reflecting its political sophistication. Agents interact over realistic social network structures, receiving personalized feeds and dynamically updating their engagement behaviors and voting intentions. Experimental conditions replicate the informational and social mobilization treatments of the original Facebook study. Across scenarios, the simulator reproduces qualitative patterns observed in field experiments, including stronger mobilization effects under social message treatments and measurable peer spillovers. Our framework provides a controlled, reproducible environment for testing counterfactual designs and sensitivity analyses in political mobilization research, offering a bridge between high-validity field experiments and flexible computational modeling.\footnote{Code and data available at \url{https://github.com/CausalMP/LLM-SocioPol}}
}


\maketitle
\section{Introduction}
\label{sec:Intro}

\begin{figure}
    \centering
    \includegraphics[width=1\linewidth]{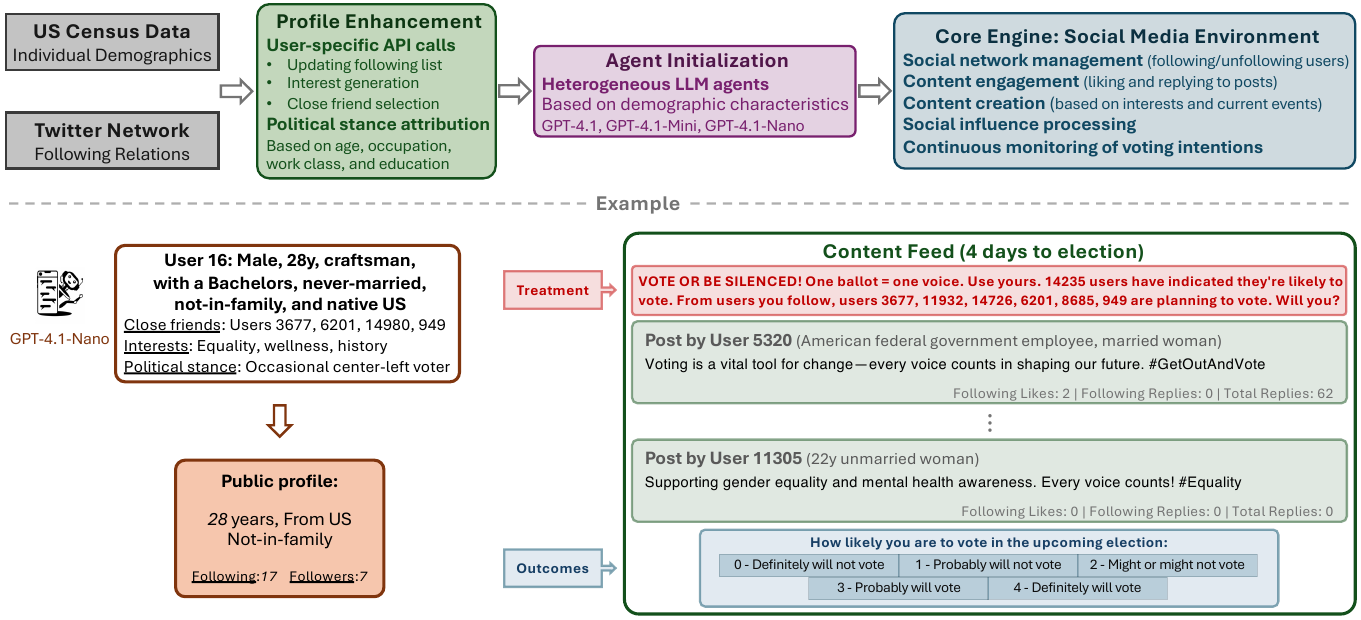}
    \caption{Overview of the LLM Social-Political Mobilization (LLM-SocioPol) simulator.
    The framework integrates real demographic and network data with heterogeneous LLM agents to model online voter mobilization. U.S. Census and Twitter data are first used to construct a realistic population and social graph. Each agent’s profile is then enhanced with demographic attributes and political-stance scores before being assigned an LLM model tier (GPT-4.1, GPT-4.1-Mini, or GPT-4.1-Nano) reflecting individual sophistication. Within the simulation environment, agents manage their follow relationships, engage with and create posts, process social-influence cues, and continuously update their voting intentions. The example panel shows a representative user profile, a treated feed containing a social-message prompt, and the 0–4 voting-likelihood scale for outcome measurement.}
    \label{fig:llm_elec}
\end{figure}

Understanding how social influence shapes political participation is a central question in political science and computational social science \citep{lazarsfeld1968people, huckfeldt1995citizens, lazer2009computational, centola2010spread}.
While face-to-face interactions have long been recognized as key drivers of voter mobilization and political engagement \citep{nickerson2008is}, the rapid expansion of online social networks has introduced new pathways through which influence can spread. Platforms such as Facebook and X (formerly Twitter) connect individuals through large-scale networks that combine strong and weak ties, allowing political messages to circulate at unprecedented scale and speed \citep{bakshy2012role}. However, the mechanisms connecting online interaction to offline electoral behavior remain only partially understood. This is mainly due to the
difficulty of identifying and measuring network-based spillover effects in real-world settings \citep{aral2009distinguishing, imbens2024causal}.

A large-scale experiment by \citet{bond201261} provides causal evidence that online interactions can shape offline political behavior. The study shows that even a single mobilization message delivered to Facebook users on election day can influence both self-reported voting behavior and verified turnout. It further finds that social messages (displaying which friends have voted) produce stronger mobilizing effects than purely informational messages, with observable spillovers to friends and friends-of-friends. These results suggest that online mobilization campaigns can use social contagion, particularly through strong-tie relationships, to amplify political participation.

Agent-based modeling (ABM) provides a complementary experimental framework to study social influence under controlled yet flexible conditions.
By constructing virtual societies in which individuals follow specified behavioral rules, ABMs allow researchers to manipulate interventions, network structures, and communication dynamics while preserving the interpretability \citep{epstein1996growing}.
When these behavioral rules reflect realistic aspects of human decision-making, iterative simulations can reveal how individual choices aggregate into collective outcomes, uncovering mechanisms that remain hidden in empirical data \citep{macy2002factors}.

Recent advances in computational social science have improved the realism of ABMs by grounding them in empirical network data and large-scale behavioral evidence \citep{lazer2020computational}.
Large language models (LLMs) extend this approach by enabling agents to make context-dependent decisions through natural language reasoning and learned social knowledge \citep{argyle2023out, park2023generative,horton2023large,gao2024large, anthis2025llm}.
These agents can interpret messages, update beliefs, and adapt their actions to changing social environments.
By combining empirical networks with cognitively rich agents, researchers can repeat the same simulated society under alternative scenarios and test counterfactual interventions. This setup enables the study of peer influence with a level of realism that was previously unattainable outside real social platforms.

In this study, we present the \textbf{LLM Social-Political Mobilization (LLM-SocioPol)} simulator, an LLM-driven agent-based social media environment designed to replicate and extend the experimental framework of \citet{bond201261}. The simulation integrates four key components:
\begin{enumerate}
    \item \textbf{Realistic population modeling}: Agents are initialized with demographic attributes sampled from U.S. Census data, ensuring representativeness in age, gender, race, education, and occupation \citep{kohavi1994data}.
    \item \textbf{Authentic network topology}: The social graph is derived from Twitter follower--following relationships, capturing realistic network structures \citep{leskovec2012learning}.
    \item \textbf{Heterogeneous, politically profiled LLM agents}: Agents’ political stances and turnout propensities are assigned algorithmically based on demographic profiles, occupation, and expressed interests, yielding a spectrum of voter types from ``consistent progressive voters'' to ``low-turnout conservatives.''
    \item \textbf{Realistic interaction dynamics}: Agents interact with the platform through behaviors that mirror real-world social media use. In each simulation round, they can like or reply to posts, create new posts inspired by their content feed, follow or unfollow other users, and decide their next login time endogenously based on engagement and their own availability. These actions drive a realistic network evolution and content flow.
\end{enumerate}

The simulation engine (Figure~\ref{fig:llm_elec}) operates over discrete time rounds leading up to election day. In each round, active agents receive personalized content feeds generated from posts in their network, engage with selected posts via likes or replies, and update their voting likelihood on a $0$--$4$ scale. Agents are heterogeneous not only in their demographic and political profiles but also in their cognitive sophistication: each agent is instantiated with one of several LLM variants (\texttt{GPT-4.1}, \texttt{GPT-4.1-Mini}, or \texttt{GPT-4.1-Nano}) according to demographic attributes such as education level and occupation. This stratification reflects empirical evidence that information processing depth and political knowledge vary systematically across the population, allowing the simulation to model heterogeneous susceptibility to mobilization and peer influence.

Treatment assignment follows the \citet{bond201261} protocol: a control group (no mobilization), an informational treatment (generic get-out-the-vote message), or a social treatment (message displaying how many users and which followed users have indicated voting intentions). The social treatment leverages strong- and weak-tie cues in the network to replicate peer influence mechanisms. On election day, a final LLM-based decision process determines whether each agent votes, integrating prior voting likelihood trajectories, treatment exposure, political stance, and demographics. This layered architecture allows the simulation to capture both micro-level heterogeneity in decision-making and macro-level patterns of turnout emerging from networked interactions.

Across five independent iterations (parallelized across 128 CPU cores and totaling about 19,200 CPU-hours with over 3 million OpenAI API calls) the simulator reproduces key qualitative findings from the original Facebook experiment. Specifically, social messages yield higher voting turnout than informational messages, and peer visibility amplifies mobilization through indirect influence, when untreated users are affected by their friends who received the message. Quantitatively, however, the observed effects differ from those in the real experiment. This is likely due to the limited social depth and the absence of offline interactions or real-world distractions in the simulated environment. These results highlight both the fidelity and the current limitations of LLM-based social simulations in capturing collective social behavior.

Overall, LLM-SocioPol recreates a realistic online mobilization environment where both direct and spillover effects can be observed. The resulting simulation outputs include detailed, round-by-round records of agent states, network exposures, content interactions, and decision trajectories. With this design, LLM-SocioPol functions as an ex-ante testbed for studying how message framing, network structure, and tie strength influence mobilization outcomes. By combining empirical insights from online experiments and recent progress in LLM-augmented agent-based modeling, the framework bridges the rigor of randomized field trials with the flexibility of simulation. This enables researchers to replicate established findings, validate causal estimators, and experiment with intervention designs that would be infeasible or ethically constrained in real-world platforms.

The rest of this article is organized as follows. We begin by reviewing related literature in Section~\ref{sec:lit_rev}. Section~\ref{sec:simulator} describes the design of the LLM-SocioPol simulator, including data sources and agent initialization, while Section~\ref{sec:exp_scenario} details the experimental scenarios. Section~\ref{sec:results} presents the main findings on voting intentions and turnout under different treatment conditions. Finally, Section~\ref{sec:conclusion} compares these results with real-world mobilization experiments and discusses their implications.

\section{Relevant Literature}
\label{sec:lit_rev}

Understanding how social influence shapes individual behavior has long been a central question in social science. Social media has provided a natural channel for such influences to spread in recent years. The Facebook voting experiment shows that information about peers’ voting behavior propagated through online connections and measurably increased turnout \citep{bond201261}. Other studies on online platforms confirmed that influence frequently flows through weak or indirect ties, amplifying behavioral diffusion beyond close relationships \citep{bakshy2011everyone,aral2011creating,rajkumar2022causal}. Collectively, this literature shows how micro-level social exposure can generate macro-level behavioral shifts. Our work builds on these insights by constructing a controlled, data-driven simulation environment that reproduces such network contagion processes.

A central challenge in these settings is designing and analyzing experiments when the independence of experimental units does not hold. Interactions among users imply that one individual’s treatment can influence others’ behavior and outcomes, a phenomenon known as \emph{network interference}. This challenge has motivated an extensive literature on causal inference under interference \citep{hudgens2008toward,manski2013identification,toulis2013estimation,ugander2013graph,eckles2016design,aronow2017estimating,athey2018exact,holtz2020reducing,hu2022switchback,li2022random,leung2022causal,farias2022markovian,yu2022estimating,bojinov2023design,shirani2024causal,johari2024does,bayati2024higher}. These studies develop estimators and design strategies that improve identification and precision when interference is present. Yet, validating these methods remains difficult because real-world experiments provide only a single realization of the experimental setting, leaving the ground-truth value of desired estimands unobservable \citep{holland1986statistics}. Following \cite{shirani2025can}, this work contributes to this literature by developing a high-fidelity simulation framework that reproduces network interference in a realistic environment, enabling systematic evaluation of experimental designs and estimators.

Traditional agent-based models have long examined diffusion and coordination in networks \citep{axelrod1997dissemination,macy2002factors}, but their behavioral rules are often stylized. Recent advances in large language models have enabled \emph{LLM-based social simulations} that exhibit context-dependent behaviors across economic, political, and online-platform environments \citep{zhao2023competeai,chuang2023simulating,park2024generative,tang2024gensim,xiao2024tradingagents,hu2024llm,ferraro2024agent,piao2025agentsociety,rehan2025synthetic,bertolotti2025llm,piao2025exploring,fontana2025nicer,wang2025agenta,composta2025simulating,cau2025language,cai2025simulation,zhang2025llm,wang2025yulan}. Generative agents exhibit humanlike interactions in persistent environments, and recent syntheses summarize rapid progress and open challenges \citep{park2023generative,anthis2025llm}. At the same time, audits and reviews document limits, especially diversity, bias, and distributional mismatch \citep{santurkar2023whose,gallegos2024bias,anthis2025llm}. Without grounding in population data, persona-based simulations may misrepresent demographic patterns \citep{li2025llm,liu2024evaluating}. Collectively, these studies highlight the importance of socially grounded, data-driven simulation frameworks. This study contributes by embedding demographic realism, real-world network structure, and experimental control into a unified modeling environment.

\section{LLM Social-Political Mobilization (LLM-SocioPol)}
\label{sec:simulator}

LLM-SocioPol is an \emph{agent-based social media environment} that replicates and extends the experimental structure of \citet{bond201261}. The goal is to enable systematic variation in treatment conditions, network structures, and agent decision rules. Building on recent advances in large language model–based social network simulations \citep{chang2024llms}, LLM-SocioPol employs heterogeneous LLM agents to represent diverse voter profiles and interaction patterns. This setup enables controlled evaluation of social influence mechanisms at scale and allows the same experiment to be repeated under alternative treatment scenarios. Each simulation proceeds in discrete rounds corresponding to days in the pre-election period and concludes with an election-day decision phase. The following sections detail the main stages of the simulation pipeline.

\subsection{Population Initialization}
In LLM-SocioPol, a \emph{user} represents an individual registered in the simulated social network. Each user is characterized by two components: a set of personal characteristics, referred to as the \emph{profile}, and a set of social connections, referred to as the \emph{network}, defined by the list of other users they follow. Every user is then modeled by a distinct LLM agent intended to capture their behavior and decision-making. To ensure realism, we construct the profile dataset from U.S. Census data \citep{kohavi1994data}, which includes demographic attributes such as age, work class, education, marital status, occupation, relationship, race, sex, weekly working hours, and native country. The network dataset is also drawn from a real-world Twitter follower network \citep{leskovec2012learning}, providing a realistic structure of social ties.

We initialize 20{,}000 users by randomly matching each demographic profile from the Census dataset to a node in the Twitter network. In real-world social media, however, users typically have only partial visibility into others’ personal details. To reflect this, we define for each user a \emph{public profile} containing a limited subset of demographic attributes alongside a unique user identifier~(ID). Whenever a user encounters another in LLM-SocioPol, only this public profile is displayed.

Across five independent iterations—parallelized across 128 CPU cores and totaling about 19,200 CPU-hours with over 3 million OpenAI API calls—the simulator reproduces key qualitative findings from the original Facebook experiment. Since the profile and network datasets originate from independent sources, random matching can produce unrealistic combinations (e.g., a highly social profile assigned to a sparsely connected node). To address this and complete agent initialization, we ask each user to update their connections and augment them with three additional attributes: close friends, interests, and political stance. These additional steps allow users both to revise their follow relationships and to enrich their profiles with personalized attributes, ensuring that social ties and interests jointly reflect realistic patterns of social alignment.

First, we update the users' networks and create their close friends list and interests through user-specific API calls to \texttt{GPT-4.1-Nano}. Each prompt includes the user’s demographic profile to establish their persona, the list of users they currently follow, a random sample of non-followed users drawn from the initial network, and a comprehensive list of possible interests. The agent is then instructed to: (i) optionally follow or unfollow users based on its own personality and other users' characteristics (which is determined by their public profile), (ii) select two to four topics of interests, and (iii) designate one to seven \emph{close friends} by announcing their ID numbers. In this context, close friends are defined informally to contrast strong ties with weak ties, placing all individuals into one of three categories: not followed, followed, or close~friend.

We then assign each user a \emph{political stance} attribute that captures both ideological orientation and baseline turnout propensity. This step is designed to ensure consistent political behavior over time despite inherent stochasticity in LLM outputs. To generate stances, we use a second AI model (\texttt{Claude Sonnet 4}), which classifies users based on their demographics and by considering realistic patterns of political opinion in the real world. Specifically, this procedure maps users into categories such as ``consistent progressive voter,'' ``moderate swing voter,'' or ``conservative-leaning low-turnout,'' combining ideological commitment with likelihood of electoral participation.

Finally, we assign each user to one of three LLM configurations: \texttt{GPT-4.1}, \texttt{GPT-4.1-Mini}, or \texttt{GPT-4.1-Nano}, based on demographic proxies for political sophistication. Users with higher education levels and cognitively demanding occupations are assigned more capable models, enabling richer reasoning and more nuanced decision-making in subsequent simulation stages.

\subsection{Simulator Prompt}
\label{sec:sim_message}
In a real-world social network, not all users are considered continuously present on the platform. Accordingly, in LLM-SocioPol, we define an \emph{activity session} as a round in which an agent becomes active, analogous to a real-world user checking their social media feed. An activity session encapsulates the complete interaction cycle for that round: the agent logs in, receives or recalls contextual information, observes and engages with content, updates their voting intention, and determines the timing of their next activity session. This framework mirrors observed patterns of intermittent platform use in real-world social networks, while maintaining the modeling flexibility to capture heterogeneous behaviors across the simulated population \citep{anthis2025llm}. Below, we describe the core stages of an activity session, from persona creation and context restoration to content presentation, voting intention elicitation, and structured decision instructions (Figure~\ref{fig:sim-message-cycle}).

\subsubsection*{Persona creation.}
In LLM-SocioPol, the session for an active user begins with a persona creation stage, in which the agent is assigned a detailed role profile that mirrors a realistic social media user. The persona is then embedded into an explicit role-conditioning prompt that frames the agent’s identity for the large language model that controls its behavior; this technique is shown to increase realism and diversity in LLM social simulations by grounding responses in individualized context \citep{park2024individualized}. This setup ensures that subsequent decisions, such as content engagement, political expression, and voting intention updates, are conditioned on a coherent and humanlike identity. This enables the simulation to reflect heterogeneity and network interactions consistent with evidence from classic ABMs \citep{bonabeau2002agent}.

\subsubsection*{Context restoration}
Each session for a returning user begins with a memory reconstruction stage, where the simulator restores the agent’s recent history to preserve behavioral continuity despite the stateless nature of LLM calls. In this stage, the system compiles a concise summary of the agent’s prior state (including its last activity session, content interactions, and the most recent voting likelihood) to reestablish decision context while controlling computational overhead and token usage (the number of text units processed per model call). This reconstructed context enables the model to condition new actions on past behavior, sustaining temporal coherence and personality consistency across rounds.

Context restoration is a central challenge in multi-agent LLM simulations \citep{zhang2025survey}. Unlike humans or reinforcement-learning agents with persistent states, LLMs forget prior interactions unless context is explicitly reintroduced. Recent frameworks such as \citet{park2024generative,argyle2023out,goodyear2025effect} address this by developing memory buffers and retrieval-based summaries to maintain long-term coherence. Building on these advances, our approach implements a lightweight and scalable reconstruction mechanism that preserves continuity at minimal cost. Without such restoration, agents lose situational memory, eroding the realism of longitudinal social interactions and weakening the credibility of simulated behavioral dynamics.

\subsubsection*{Content feed.}
At the start of each session, the system assembles a personalized feed of up to five posts for the active agent. Each post contains the author’s text plus lightweight context so the agent can make realistic engagement choices (like, reply, or skip): the author’s ID and public profile, whether the agent follows the author, counts of likes and replies from the agent’s followed users, the total reply count, and one representative reply to anchor discussion (Figure~\ref{fig:llm_elec}).

Including this lightweight context into the posts serves a critical psychological and social function. In real social media environments, users rarely evaluate messages in isolation: the identity of the author and visible cues from their network shape how content is interpreted and acted upon. Displaying limited information such as a username, demographic cues, and engagement signals allows LLM agents to differentiate between familiar and unfamiliar users and infer social closeness. This design deepens the realism of the simulated network by anchoring each post in a recognizable social identity, enabling more nuanced, relationship-dependent engagement patterns. Without such contextual framing, all posts would appear homogeneous, flattening social differentiation and weakening the model’s ability to reproduce authentic peer influence dynamics.

We build the feed in three stages. First, an eligibility pass selects ``fresh'' posts that the agent has not seen or that have received new replies since the last view. Second, we personalize by bucketing candidates into (i) interest-matched posts, (ii) posts from followed authors, (iii) trending items with high total engagement (likes + replies), and (iv) an exploration pool; we sample from these buckets to meet target mix ratios and backfill to five items. Third, a feed-ranking step orders the list, prioritizing (1) posts from followed users, (2) posts with engagement by followed users, and (3) posts with higher total engagement.

\subsubsection*{Voting intention survey.}
Following feed presentation, the agent is prompted with a time-to-election statement (e.g., ``The election is in 4 days'') and asked to report its likelihood of voting on a discrete five-point scale ranging from 0 (``definitely will not vote'') to 4 (``definitely will vote''). The prompt is framed in relation to the agent’s demographic and attitudinal profile to maintain consistency with its assigned persona. This procedure aims to produce a consistent standardized measure of self-reported voting intention, which is recorded at each activity session. This consistency enables rigorous longitudinal analysis of how voting intentions change as the election approaches and allows the design of experimental scenarios to evaluate turnout propensity.

\subsubsection*{Guideline instructions.}
This stage is the decision instruction phase, in which the system provides the agent with explicit guidelines for generating its engagement and state update for a subsequent step. After the curated and ranked posts are presented, the agent receives detailed instructions to produce a machine-readable JSON object conforming to a fixed schema. For each post, the schema requires the engagement type (``nothing,'' ``like,'' or ``reply''), an optional reply text if ``reply'' is chosen, and a follow/unfollow decision for the post author. The output must also include an updated voting likelihood on a 0–4 scale, anchored to the agent’s political stance, and a next activity time indicating the gap until the agent's intention for the next simulated login.

The instructions enforce strict JSON formatting to enable automated parsing, prohibit extraneous output, and constrain engagement patterns to reflect realistic social media behavior. For example, agents are instructed to \emph{prefer} “likes” over “replies,” using replies sparingly. Engagement likelihood increases for posts that appear earlier in the feed, align with the agent’s values, are authored by followed users, or have engagement from followed users. The inclusion of prompts about the next activity time models the realistic tendency for users who enjoy content or have more available time to log in more frequently. Overall, this stage centers on delivering the structured prompt; the agent’s JSON-formatted response is then generated and recorded in the subsequent~stage.

\subsubsection*{Optional content creation.}
In addition to engagement, the simulation introduces a controlled level of content creation to sustain a dynamic information environment. Specifically, during each activity session, a fixed fraction of active agents are randomly selected to generate new posts, preventing feed saturation or depletion. Each selected agent is prompted to write a post that reflects its persona attributes, declared interests, and, when relevant, current political or social topics. The generated text is added to the global content pool and becomes available for inclusion in other agents’ feeds in subsequent rounds. Maintaining a consistent rate of content generation preserves diversity and freshness in the information stream, supporting realistic modeling of information diffusion and opinion dynamics.

\subsection{Agent Response}
\label{sec:agent_resp}
In the response stage, each LLM-SocioPol agent outputs a structured record of its actions, including content engagement, follow behavior, updated voting likelihood, and next activity time.
The standardized schema enables automated parsing and ensures consistency across rounds, allowing the simulation to track both individual responses and resulting network effects (Figure~\ref{fig:agent-message}).

\subsection{Event Loops}
\label{sec:sim_dynamics}
The simulation proceeds through a sequence of discrete event loops that govern agent activity and system state updates. At the start, each agent is assigned a random initial login time drawn from a uniform distribution, ensuring asynchronous entry into the simulated social media environment. The early phase of the simulation includes a warmup period, during which agents’ activity patterns and content flows stabilize before experimental treatments are introduced. This prevents transient initialization effects from biasing experimental outcomes.

During each event loop, the simulator identifies agents whose activity time has elapsed. These agents receive their updated feed (as outlined in \S~\ref{sec:sim_message}) and return structured responses (as described in \S~\ref{sec:agent_resp}). At the end of each round, the system updates four key states as follows.

\subsubsection*{Agent state:}
Internal attributes of each active agent are updated after processing its interaction with the simulator. This includes revised voting likelihoods, such as moving from “might or might not vote” to “probably will vote” in response to persuasive content or peer activity.

\subsubsection*{Network topology:}
The structure of the social graph is modified to reflect follow/unfollow actions by agents. These updates are behaviorally grounded: if an agent sees a post from a user they follow and finds it disagreeable, they may unfollow that user; conversely, if they encounter a compelling or personally relevant post from someone they do not follow, they may choose to follow the author. This mechanism allows ties to strengthen or weaken in direct response to content interactions, mirroring how real-world social networks evolve through ongoing exposures.

\subsubsection*{Content state:}
The set of posts available in the system is updated to include newly authored messages. These posts can range from non-political updates to election-related advocacy, such as calls for voter registration. The system retains metadata on who created each post, who interacted with it, and how widely it has circulated, tracking the stream of influence and reach.

\subsubsection*{Scheduling state:}
After the first login (assigned at random), the scheduling of each agent’s future participation is determined by its \emph{real next-activity-time}:
computed as the sum of the agent’s self-reported next-activity-time from its last response and an independent random perturbation.
$$
\texttt{real next-activity-time}
=
\texttt{self-reported next-activity-time}
+
\texttt{random perturbation}.
$$
This adjustment introduces stochastic variability in activity, capturing unexpected fluctuations in individual availability and attention.

This event-driven structure allows the simulation to evolve endogenously, with dynamics driven by agents’ own behaviors rather than by a fixed-step clock or externally imposed updates. As activity patterns, content flows, and network ties change over time, agents experience varying levels of peer influence and interaction. These mechanisms reproduce the irregular participation and diverse engagement rhythms observed on real social platforms. The simulation concludes with the election-day stage, where final voting decisions are recorded.

\subsection{Election Day and Voting Inquiry}
Finally, on election day, all eligible agents enter the voting inquiry stage, which determines their final participation outcome. Each agent receives a structured prompt summarizing their demographic profile, political stance, most recent voting likelihood, and the decision rule for responding. The inquiry is deliberately minimalistic to avoid introducing new persuasive content at this final stage. Agents respond with a binary value—1 if they vote or 0 if they abstain. This outcome reflects the cumulative effects of demographic predispositions, peer influence, treatment exposure, and prior engagement throughout the simulation. These binary voting decisions are then recorded as the primary endpoint for turnout analysis (Figure~\ref{fig:voting-message}).

\begin{table}[t]
\footnotesize
\setlength{\tabcolsep}{3pt}
\renewcommand{\arraystretch}{1.05}
\centering
\begin{tabularx}{\linewidth}{@{} l Y l @{}}
\toprule
\textbf{Component} & \textbf{Compact description} & \textbf{Source / Updated by} \\
\midrule
Profile (full) & Demographics: age, gender, race, etc. & U.S. Census; fixed \\
Public profile & Subset of profile & Derived by system; fixed \\
Network (following) & Directed follows for each user & Twitter graph; updated by initial API call \\
Close friends & 1–7 designated strong ties & Initial API call; fixed \\
Interests & 2–4 topical interests & Initial API call; fixed \\
Political stance & Ideology + baseline turnout propensity & External model (\texttt{Claude Sonnet 4}); fixed \\
LLM config & \texttt{GPT-4.1}/\texttt{4.1-Mini}/\texttt{4.1-Nano} per user & System assignment; fixed \\
Feed (per session) & Personalized list of up to 5 posts & System ranking from network and content \\
Agent decision & Engagement to posts, follow/unfollow actions, voting likelihood, next-activity-time & LLM agent; tracked over activity sessions \\
Content pool & All posts available to feeds & Agents generate; grows over time \\
Real next-activity-time & self-reported + random noise & Agent sets base; system adds perturbation \\
Voting likelihood (0–4) & Stated propensity to vote & LLM agent; tracked over activity sessions \\
Election-day vote & Final turnout (0/1) & LLM agent; generated at the end \\
Treatment scenario & Control/informational/social & Randomized by experimenter \\
\bottomrule
\end{tabularx}
\caption{Core components of \textit{LLM-SocioPol}. Column 3 indicates whether a component is fixed (exogenously initialized) or updates endogenously via agent choices (and, where relevant, system stochasticity).}
\label{tab:components}
\end{table}

\section{Experimental Setting}
\label{sec:exp_scenario}
In LLM-SocioPol, we define multiple outcome measures to analyze the effects of different interventions on voter mobilization and behavior over time. Following \citet{bond201261}, we consider each agent’s \emph{election-day vote} as the primary outcome, recorded once on the final day of the simulation. Extending that framework, we also track each agent’s \emph{voting likelihood} (on a 0–4 scale) across pre-election days as a temporal outcome, observed whenever the agent becomes active. These repeated measures capture how agents’ voting intentions evolve under different treatments, allowing us to quantify changes in engagement and mobilization dynamics as the election day approaches.

We also consider three default experimental scenarios within the LLM-SocioPol framework to examine how alternative voting campaign messages influence electoral participation. Only agents representing voting-age individuals (18 years or older) are eligible for random assignment. In all scenarios, agents view the treatment message as part of their voting-intention survey (Figures~\ref{fig:sim-message-cycle} and~\ref{fig:treatment-message}).
This ensures that exposure occurs immediately before they report their likelihood of voting.
This design follows the approach of \citet{bond201261}, where survey prompts were used to deliver mobilization messages. We explain each scenario below.

\subsubsection*{Control (no additional message).}
Agents receive the baseline voting intention survey with no mobilization message. This condition serves as the reference point for measuring treatment-induced changes, capturing natural variation in voting likelihood absent any explicit campaign stimulus.

\subsubsection*{Informational message.}
Treated agents in this scenario see a banner above the survey stating:
\begin{quote}
\small\texttt{``VOTE OR BE SILENCED! One ballot = one voice. Use yours.''}
\end{quote}
This condition parallels the informational treatment in \citet{bond201261}, which provided an election-day reminder and motivational framing without any peer-specific cues. In the original study, the message also included polling place information; in LLM-SocioPol, location-specific details are omitted because voting location is not modeled as a logistical constraint. This scenario tests whether factual mobilization appeals, absent social proof, are sufficient to change intentions.

\subsubsection*{Social message.}
In this scenario, treated agents see the same informational banner, followed by a counter of how many users have stated they will vote, and a list of followed users who have already indicated their voting intention:
\begin{quote}
\small\texttt{``VOTE OR BE SILENCED! One ballot = one voice. Use yours. 14625 users have indicated they're likely to vote. From users you follow, users 3677, 12604, 11932, 16785, 949, 14726 are planning to vote. Will you?''}
\end{quote}
This scenario parallels the social treatment in \citet{bond201261}, where identifiable peer actions and real-time participation counts aim to increase mobilization through social contagion. In the original Facebook experiment, peer activity was shown by displaying profile pictures of friends who had clicked the “I Voted” button, reinforcing participation through both visual and textual cues. In LLM-SocioPol, we replace profile photos with usernames of up to six followed users who have indicated their intention to vote. Although this omits the visual cue, explicitly naming known peers still provides a personalized and socially salient prompt. Such social references are among the most effective predictors of political participation in both offline canvassing and online mobilization, as they personalize the message and emphasize prevailing social norms within each agent’s immediate network \citep{nickerson2008is, gerber2008social, bond201261, bakshy2015exposure}.

By holding demographic profiles, network topology, and feed ranking algorithm constant across scenarios, the experiment isolates the causal effect of message type on both stated voting likelihood and simulated turnout. This design enables a clean comparison across four mechanisms:
(i) pure informational effects,
(ii) weak-tie peer influence (exposure through followed users),
(iii) strong-tie peer influence (exposure through close friends, which is also provided to the agent, see Figure~\ref{fig:llm_elec}), and
(iv) large-scale social norm impact.
These contrasts parallel the decomposition in \citet{bond201261} while mirroring it through a controlled, agent-based simulation environment.

\section{Results}
\label{sec:results}
We implement LLM-SocioPol through five independent iterations, each starting with 15 warm-up rounds without treatment. The simulation then unfolds dynamically, allowing agents’ exposures, interactions, and behavioral responses to evolve over successive rounds. In every iteration, we evaluate five treatment scenarios:
\begin{enumerate}
    \item ``Control'': all users remain untreated,
    \item ``Info'': all users receive the informational message,
    \item ``Social'': all users receive the social message,
    \item ``Exp Info'': the informational message is introduced gradually to 20\%, 40\%, and 80\% of users across three consecutive 10-round blocks, following a staggered rollout design, and
    \item ``Exp Social'': the same staggered rollout applied to the social message.
\end{enumerate}
In the staggered rollout design, treatment is introduced sequentially; therefore, users treated in earlier stages remain exposed for the rest of the experiment, while new users are added in subsequent phases \citep{xiong2024optimal}. Across iterations, we vary the random seed controlling LLM model assignment (across \texttt{GPT-4.1} variants), user activity timing (the random perturbation component), and treatment randomization.

Five simulation iterations are executed using OpenAI API on the Stanford Graduate School of Business Data, Analytics, and Research Computing Core Facility (RRID:SCR\_022938). Each implementation is parallelized across 128 CPU cores and requires approximately 30 hours of wall-clock time. A single iteration, consisting of five full runs (one for each treatment scenario), therefore consumes about 19,200 CPU hours. Each round triggers one LLM inference per active agent, leading to more than 600,000 API calls per iteration and over 3 million across all runs, depending on agents’ endogenous activity patterns. Below, we summarize our observations from LLM-SocioPol and compare them with the findings of the real-world experiment reported in \citet{bond201261}.

\subsection{Voting Turnout}
Our main objective is to validate the central finding of \citet{bond201261}: social messages show a significant positive effect on voter turnout, whereas informational messages have no measurable impact. To assess this pattern, we first use the binary voting outcomes recorded on the simulated election day and estimate treatment effects under different experimental conditions. Specifically, we employ the Difference-in-Means (DM) estimator \citep{imbens2015causal}:
\begin{align}
    \hat{\tau}_{\text{DM}} = \frac{1}{N_T}\sum_{i=1}^{N} Y_i \cdot W_i - \frac{1}{N_C}\sum_{i=1}^{N} Y_i \cdot (1-W_i),
\end{align}
where $Y_i$ is the binary voting outcome for agent $i$, $W_i$ is the treatment indicator (1 if treated, 0 if control), and $N_T$ and $N_C$ are the number of treated and control agents, respectively. The standard error for the DM estimator is calculated as $\text{SE}_{\text{DM}} = \sqrt{\frac{\sigma_1^2}{N_T} + \frac{\sigma_0^2}{N_C}}$, with $\sigma_1^2$ and $\sigma_0^2$ denoting the sample variances within treatment and control groups. We also compute “ground-truth” treatment effects from the full-treatment runs (where all agents receive the intervention).

\begin{figure}[ht]
\centering
\includegraphics[width=0.9\textwidth]{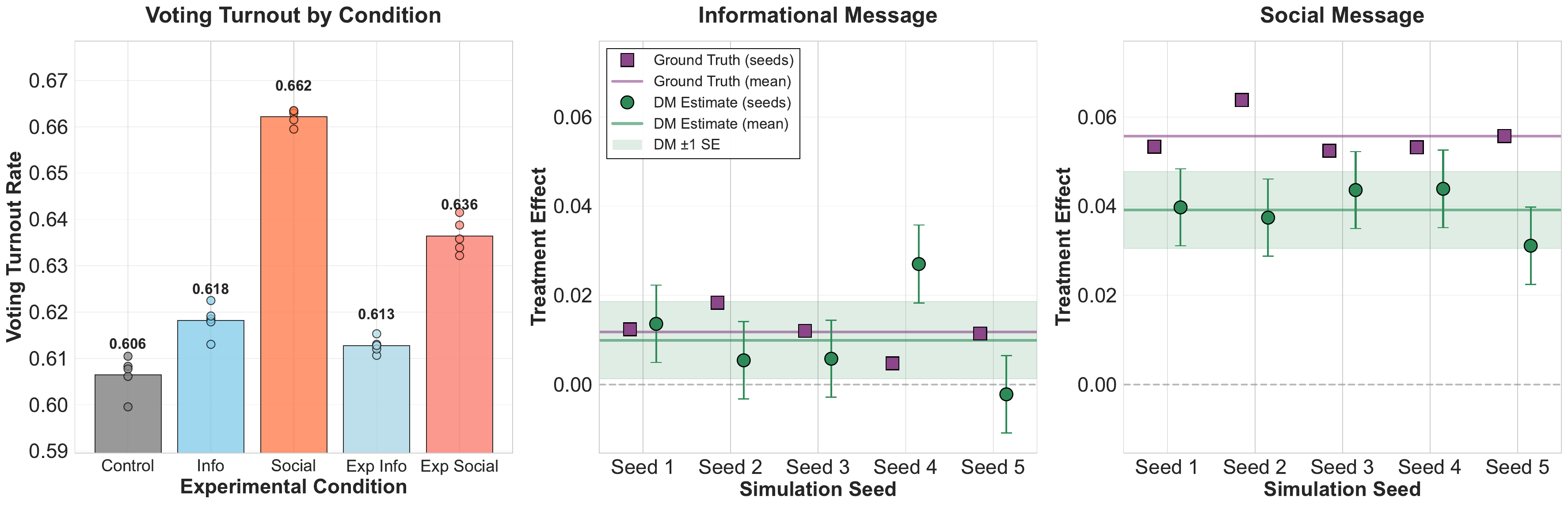}
\caption{Voting turnout and Difference-in-Means (DM) estimates across experimental conditions.
Left: Average turnout across five random seeds (19,785 agents each; hollow circles show individual seeds).
Middle and right: Estimated effects of informational and social messages on voting turnout. Social messages boost turnout far more than informational ones, mirroring the qualitative pattern in \citet{bond201261}.}
\label{fig:turnout_analysis}
\end{figure}

LLM-SocioPol reproduces the qualitative findings of \citet{bond201261}. In the original study involving 61 million users, social messages that display peers’ voting behavior increased voting turnout relative to the control, while informational messages produced no detectable effect. Applying the DM estimator to our experimental scenarios yields consistent results: the estimator reliably identifies a positive effect of social messages on turnout, whereas informational messages show no significant effect in three of the five simulation iterations (Figure~\ref{fig:turnout_analysis}).

When comparing the all-control and all-treatment scenarios, our simulator exhibits a similar pattern with amplified magnitudes. On average, social messages raise turnout by 5.6\% across five runs, while informational messages increase turnout by 1.2\%. Although the experimental estimator does not always detect this smaller informational effect, the ground-truth simulations suggest that low-signal effects of purely informational exposure may exist. However, these effects are difficult to detect empirically because of their small magnitude.

\begin{figure}[ht]
    \centering
    \includegraphics[width=0.9\textwidth]{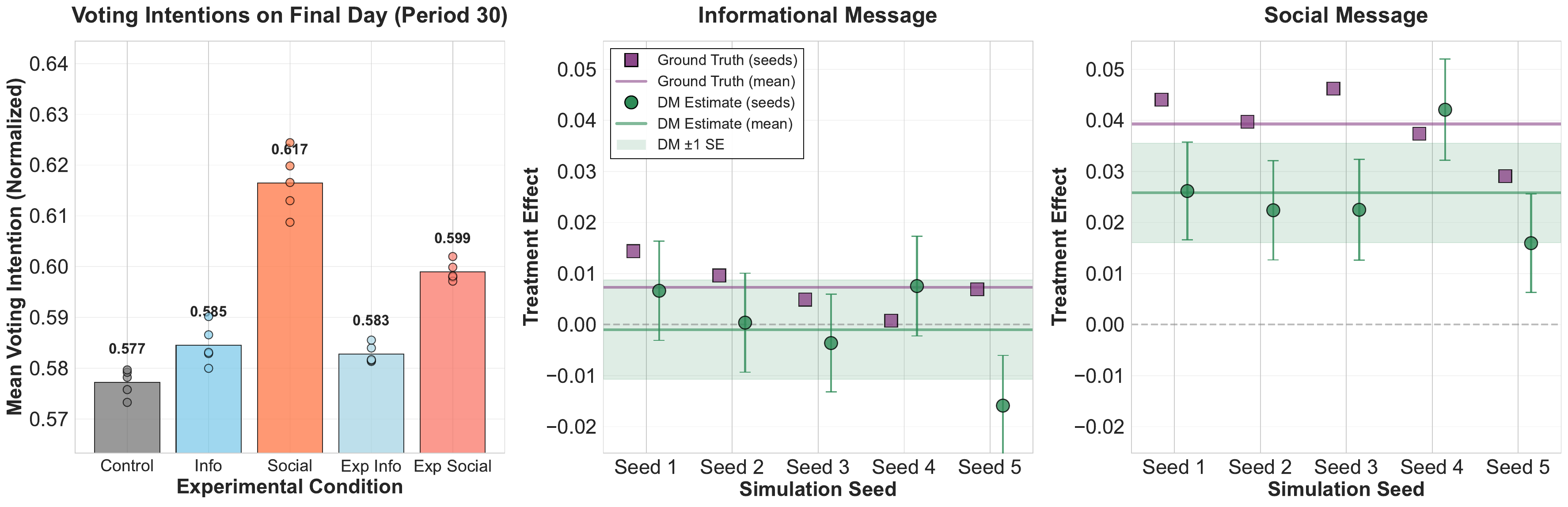}
    \caption{Normalized voting intentions (to 0--1 scale) on final pre-election period (period 30). Left: mean intentions. Middle and right: Estimated effects of informational and social messages on voting intention.}
    \label{fig:final_intentions}
\end{figure}

\subsection{Voting Intentions}
\begin{figure}[ht]
    \centering
    \includegraphics[width=0.7\textwidth]{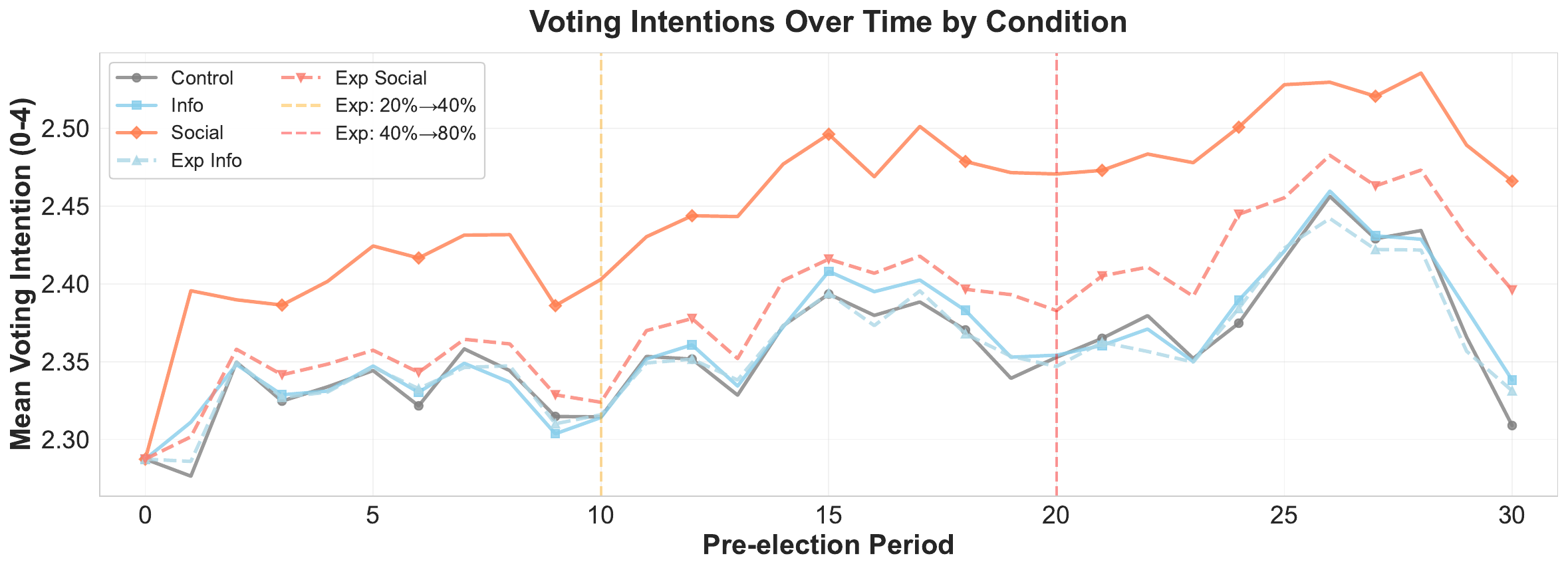}
    \caption{Mean voting intentions (0--4 scale) over pre-election periods. Vertical lines mark treatment phase transitions. The social message consistently demonstrates higher intentions than the informational message.}
    \label{fig:intentions_timeseries}
\end{figure}

We extend the turnout analysis by tracking voting intentions over the 31 pre-election periods. Figure~\ref{fig:intentions_timeseries} displays the evolution of mean intentions (0–4 scale) across all five conditions. The social message treatment (both in the full and experimental scenarios) consistently produces higher intentions than the informational message, echoing the turnout results in Figure~\ref{fig:turnout_analysis}.

In the Control, Info, and Social scenarios where all users receive the same treatment, intentions generally rise as election day approaches, likely reflecting increasing campaign salience and peer activity within the simulated network. A sharp decline then appears in the final days, which may represent a stabilization phase where voters consolidate their decisions and engagement intensity subsides. Further work is needed to assess whether this late drop originates from simulation dynamics or mirrors real-world mobilization behavior.

Building on the expectation that final turnout should reflect the most recent reported intentions, we focus on the last pre-election period. Figure~\ref{fig:final_intentions} reports normalized voting intentions (0–1 scale). Across all conditions and seeds, between 3,719 and 3,898 agents were active before election day. Among these users, mean intentions increase notably under the social message treatment but remain close to control levels under the informational message. DM estimates align with ground truth for informational messages yet underestimate the social message effect, confirming that only social messaging produces a meaningful rise in pre-election voting intentions.

\begin{figure}[ht]
    \centering
    \includegraphics[width=\textwidth]{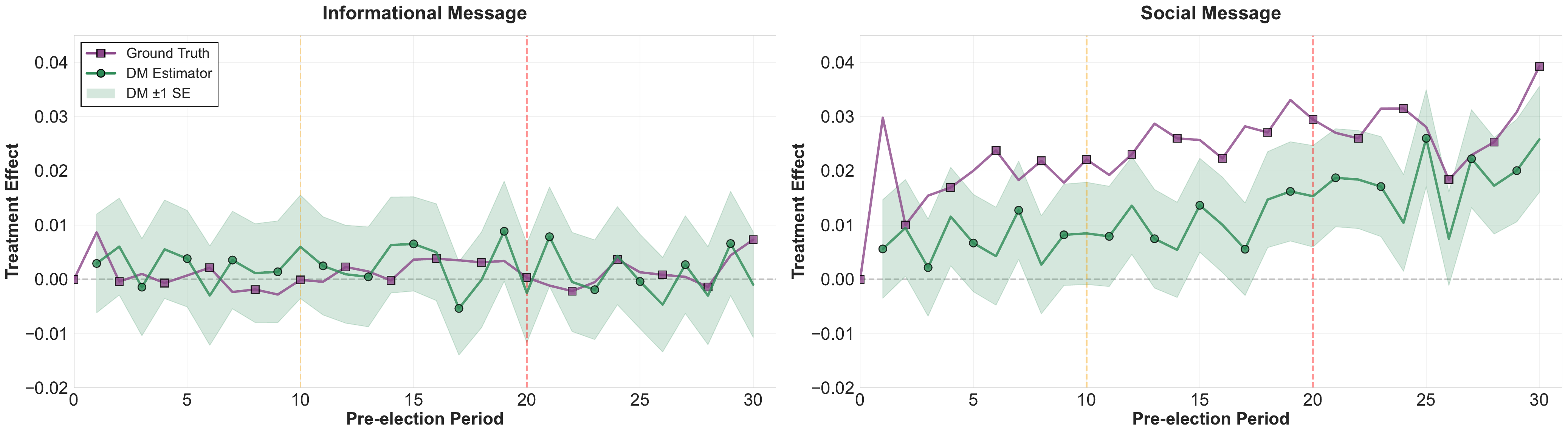}
    \caption{Treatment effects on normalized voting intentions aggregated across five seeds. Left: informational message. Right: social message. Vertical lines mark treatment phase transitions.}
    \label{fig:intentions_aggregated}
\end{figure}

Figure~\ref{fig:intentions_aggregated} traces the evolution of treatment effects on normalized voting intentions, averaged across all five simulation seeds. The informational message shows no measurable impact, with both ground truth and DM estimates fluctuating around zero throughout the campaign. In contrast, the social message generates steadily rising effects, from about 1.5 percentage points in early rounds to roughly 4 by period 30. Although the DM estimator follows the same upward pattern, it consistently underestimates the true effect. These results indicate that the influence of social messaging accumulates over time, while standard estimators systematically miss part of the effect.

\section{Discussion and Conclusion}
\label{sec:conclusion}
This paper presents LLM-SocioPol, an agent-based social media simulator that combines real demographic data, authentic network structure, and LLM agents to study how social influence shapes voter mobilization. The simulator reproduces the design of the 61-million-person Facebook experiment of \cite{bond201261} within a controlled, repeatable environment. Using 20,000 synthetic users, the study examined how informational and social voting messages affect both pre-election voting intentions and final turnout. By employing LLM-driven decision loops, LLM-SocioPol demonstrates how language models can replicate and extend large-scale experiments, enabling investigations of social contagion and digital interventions in networked populations.

The simulation results (Figures~\ref{fig:turnout_analysis}–\ref{fig:intentions_aggregated}) show that social messages consistently generate stronger mobilization than informational messages, mirroring the real-world findings that peer cues amplify participation. Because the difference-in-means estimator in network experiments captures only the direct treatment effect, excluding spillovers through social ties \citep{savje2021average,munro2025treatment}, its consistent underestimation of the true effect in our results indicates strong network interference. This pattern qualitatively reaffirms the original Facebook study’s conclusion that social influence propagates through network connections.

However, the simulated magnitudes differ from those observed in the real experiment. In our setting, the direct effect (estimated by the difference-in-means) of the social message is about 3.9\% on turnout (Figure~\ref{fig:turnout_analysis}) and 2.6\% on final voting intentions (Figure~\ref{fig:final_intentions}). The corresponding indirect effects (calculated as the difference between the ground-truth and direct effect estimates) are 1.7\% and 1.4\%, respectively. In contrast, \citet{bond201261} reported a direct effect of 0.39\% on validated turnout, while network contagion produced roughly 280,000 additional votes relative to 60,000 from direct mobilization, a ratio of about 4.7 to 1. In our simulation, the analogous ratio is closer to 0.4 to 1, indicating that network spillovers are comparatively weaker in magnitude even though the qualitative direction of effects remains consistent.

Overall, the simulated effects are an order of magnitude larger than those in \cite{bond201261}, while the ratio of network to direct effects is substantially higher in the original study. This qualitative consistency but quantitative discrepancy has several implications, which we discuss below.

First, in LLM-SocioPol, agents are exposed to campaign messages beginning 30 days before the election. This prolonged exposure allows mobilization effects to accumulate over time, as reflected in the rising voting intentions shown in Figure~\ref{fig:intentions_aggregated}. \citet{bond201261} also acknowledge that their intervention likely underestimates treatment effects, as it was limited to an election-day message. Other voter mobilization experiments \citep{nickerson2008is,gerber2008social,bryan2011motivating} show that contacting potential voters can increase turnout by approximately 1–10\%, a range consistent with the magnitudes observed in our simulation.

Second, the simulated environment of LLM-SocioPol represents an isolated world in which agents interact primarily through the modeled social network. They are not exposed to real-world distractions, competing influences, or resource constraints. As a result, treatment effects appear stronger in this controlled setting than in the complex, noisy environments faced by human users.

Third, real-world social networks have greater depth, extending beyond online interactions and peer posts. Face-to-face communication, in particular, plays an important role in political mobilization \citep{gerber2000effects}. Such offline interactions are absent in LLM-SocioPol, meaning the simulation misses many pathways of influence that operate outside social media. Consequently, the simulated network effects appear weaker in the online-only environments.

In addition, comparing simulated voting turnout (Figure~\ref{fig:turnout_analysis}) and voting intentions (Figure~\ref{fig:final_intentions}), both the qualitative patterns and quantitative magnitudes appear comparable. This suggests that agents’ stated intentions provide a reliable proxy for their eventual voting behavior. Furthermore, the temporal trend in voting intentions (Figures~\ref{fig:intentions_timeseries} and \ref{fig:intentions_aggregated}) indicates that the effects of social messages accumulate gradually rather than appearing instantaneously. This pattern is consistent with a diffusion process operating through social networks, as documented in prior studies of behavioral contagion and online influence \citep{centola2010spread,aral2012identifying,christakis2013social}.

In conclusion, existing evidence suggests that the original Facebook experiment may underestimate the full effects of social mobilization, while LLM-SocioPol tends to amplify effect sizes under controlled conditions. Nonetheless, both reveal consistent qualitative patterns, highlighting the central role of network-driven influence in voter behavior. Taken together, these findings suggest that high-fidelity simulations using large language models can complement empirical studies by helping to explore mechanisms that are difficult to isolate in real-world settings.

\begin{figure}[t]
\centering
\begin{tikzpicture}[
  font=\small,
  node distance=6mm,
  stage/.style={draw, rounded corners=2pt, fill=blue!6, align=left, inner sep=3pt, minimum width=38mm},
  arrow/.style={-stealth, thick, draw=black!60},
  every node/.style={align=left}
]

\node[draw, rounded corners=1pt, fill=gray!5, anchor=north west, inner sep=6pt, font=\ttfamily\scriptsize] (bg) at (0,0) {%
\begin{minipage}{0.97\textwidth}
\colorbox{gray!25}{%
\begin{minipage}{0.99\textwidth}
You are, as a social-network user, a 28y Black Male from United-States, full-time worker with very long hours, never-married, Bachelors, Craft-repair, interests:Racial Equality,Gender Equality,Wellness,History. You are close-friend with Users:3677,6201,14980,949. 
Round:19
\end{minipage}%
}\\[2pt]
\colorbox{gray!25}{%
\begin{minipage}{0.99\textwidth}
Last Round:13\\
Last Engagement:\\
Post: Supporting policies that promote responsible travel and climate action. Every vote counts in shaping our future.\\
Action: like\\
Post: Fighting for racial equality and justice is vital. We must stay active and push for real change. \#RacialEquality \#Justice\\
Action: like\\
Your previous voting likelihood: 2/4
\end{minipage}%
}\\[2pt]

\colorbox{gray!25}{%
\begin{minipage}{0.99\textwidth}
\textbf{Feed:}\\
Post:1\\
Author:user\_id: 19147, education: HS-grad, occupation: Protective-serv, workclass: Local-gov\\
Followed: No\\
Topic: general-History\\
Following Likes: 1\\
Following Replies: 0\\
Total Replies: 0\\
Text: Supporting progress and equality in our communities. Every voice matters! \#Unity \#Justice
\\
Post:2 ...
\\
Post:3 ...
\\
Post:4 ...
\\
Post:5 ...
\end{minipage}%
}\\[2pt]

\colorbox{gray!25}{%
\begin{minipage}{0.99\textwidth}
The election is in 11 days.\\
As someone with your background and values, consider realistically how likely you are to vote in the upcoming election:\\
On a scale from 0 to 4, where:\\
0 - Definitely will not vote\\
1 - Probably will not vote\\
2 - Might or might not vote\\
3 - Probably will vote\\
4 - Definitely will vote\\[2pt]
In your last response (round 13), your voting likelihood was: 2/4.
\end{minipage}%
}\\[2pt]

\colorbox{gray!25}{%
\begin{minipage}{0.99\textwidth}
Given above posts, decide which posts to engage with. Reply ONLY in valid JSON format, NO additional text or comments.
Output must precisely match this schema:\\
\{\\
\phantom{xx}"engagement": [\\
\phantom{xxxx}\{\\
\phantom{xxxxxx}"post\_number": <int>,\\
\phantom{xxxxxx}"engage": "nothing"|"like"|"reply",\\
\phantom{xxxxxx}"reply\_text": <string>,\\
\phantom{xxxxxx}"follow\_action": "follow"|"unfollow"|"no\_change"\\
\phantom{xxxx}\}, \ldots\\
\phantom{xx}],\\
\phantom{xx}"voting\_likelihood": <int>, // Must be 0-4 integer **based on your POLITICAL STANCE: Slight center-left, rarely votes**\\
\phantom{xx}"next\_activity\_time": <int> // Must be between 0 and 6 based on your persona and engagement\\
\}\\[1pt]
\textbf{CRITICAL RULES:}\\
1.Use DOUBLE QUOTES for ALL keys and string values\\
2.Do NOT use any escape characters like \textbackslash , \textbackslash" or \textbackslash'\\
3.Include "reply\_text" field ONLY when "engage" is "reply"\\
4.Include "follow\_action" for each post to decide whether to follow/unfollow the author, if you like to see more/less from them\\
5.Ensure ALL JSON is properly terminated with closing brackets\\
6.STRONGLY prefer "like" over "reply", Very RARELY use "reply"\\
7.Posts higher in the feed, higher post-sentiment, with engagement from users you follow, or from authors you follow \textrightarrow\ MORE LIKELY TO ENGAGE\\
8.Your response MUST be ONLY the JSON object, nothing else\\
**React as a real social media user would with your persona, political stance, social ties, and values.**
\end{minipage}%
}
\end{minipage}%
};

\end{tikzpicture}
\caption{Example of simulation dynamics for an activity session of User 16. From top to bottom: persona creation, context restoration, content feed, voting intention survey, and guideline instructions.}
\label{fig:sim-message-cycle}
\end{figure}

\begin{figure}[t]
\centering
\begin{tikzpicture}[
  font=\small,
  node distance=6mm,
  stage/.style={draw, rounded corners=2pt, fill=blue!6, align=left, inner sep=3pt, minimum width=38mm},
  arrow/.style={-stealth, thick, draw=black!60},
  every node/.style={align=left}
]

\node[draw, rounded corners=1pt, fill=gray!5, anchor=north west, inner sep=6pt, font=\ttfamily\scriptsize] (bg) at (0,0) {%
\begin{minipage}{0.97\textwidth}
\colorbox{gray!25}{%
\begin{minipage}{0.99\textwidth}
\{\\
\phantom{xx}"engagement": [\\
\phantom{xxxx}\{\\
\phantom{xxxxxx}"post\_number": 1,\\
\phantom{xxxxxx}"engage": "like",\\
\phantom{xxxxxx}"reply\_text": "",\\
\phantom{xxxxxx}"follow\_action": "no\_change"\\
\phantom{xxxx}\},\\
\phantom{xxxx}\{\\
\phantom{xxxxxx}"post\_number": 2,\\
\phantom{xxxxxx}"engage": "like",\\
\phantom{xxxxxx}"reply\_text": "",\\
\phantom{xxxxxx}"follow\_action": "no\_change"\\
\phantom{xxxx}\},\\
\phantom{xxxx}\{\\
\phantom{xxxxxx}"post\_number": 3,\\
\phantom{xxxxxx}"engage": "nothing",\\
\phantom{xxxxxx}"reply\_text": "",\\
\phantom{xxxxxx}"follow\_action": "no\_change"\\
\phantom{xxxx}\},\\
\phantom{xxxx}\{\\
\phantom{xxxxxx}"post\_number": 4,\\
\phantom{xxxxxx}"engage": "nothing",\\
\phantom{xxxxxx}"reply\_text": "",\\
\phantom{xxxxxx}"follow\_action": "no\_change"\\
\phantom{xxxx}\},\\
\phantom{xxxx}\{\\
\phantom{xxxxxx}"post\_number": 5,\\
\phantom{xxxxxx}"engage": "nothing",\\
\phantom{xxxxxx}"reply\_text": "",\\
\phantom{xxxxxx}"follow\_action": "no\_change"\\
\phantom{xxxx}\}\\
\phantom{xx}],\\
\phantom{xx}"voting\_likelihood": 2,\\
\phantom{xx}"next\_activity\_time": 3\\
\}
\end{minipage}%
}\\[2pt]

\end{minipage}%
};

\end{tikzpicture}
\caption{Example of agent's response.}
\label{fig:agent-message}
\end{figure}

\begin{figure}[t]
\centering
\begin{tikzpicture}[
  font=\small,
  node distance=6mm,
  stage/.style={draw, rounded corners=2pt, fill=blue!6, align=left, inner sep=3pt, minimum width=38mm},
  arrow/.style={-stealth, thick, draw=black!60},
  every node/.style={align=left}
]

\node[draw, rounded corners=1pt, fill=gray!5, anchor=north west, inner sep=6pt, font=\ttfamily\scriptsize] (bg) at (0,0) {%
\begin{minipage}{0.97\textwidth}
\colorbox{gray!25}{%
\begin{minipage}{0.99\textwidth}
Message 17 (Simulator):
You are, as a social-network user, a 28y Black Male from United-States, full-time worker with very long hours, never-married, Bachelors, Craft-repair, interests:Racial Equality,Gender Equality,Wellness,History. You are close-friend with Users:3677,6201,14980,949. \\

This is the election day. Your political stance is (Slight center-left, rarely votes) and your most recent voting likelihood was 2/4, where:\\
0 - Definitely will not vote\\
1 - Probably will not vote\\
2 - Might or might not vote\\
3 - Probably will vote\\
4 - Definitely will vote\\

Considering your **voting likelihood of 2/4**, reply with ONLY `1' if you voted or `0' if you did not vote.

\end{minipage}%
}\\[2pt]

\colorbox{gray!25}{%
\begin{minipage}{0.99\textwidth}

Message 18 (Agent - DID NOT VOTE):
0

\end{minipage}%
}
\end{minipage}%
};

\end{tikzpicture}
\caption{Example of voting inquiry and agent's response.}
\label{fig:voting-message}
\end{figure}

\begin{figure}[t]
\centering
\begin{tikzpicture}[
  font=\small,
  node distance=6mm,
  stage/.style={draw, rounded corners=2pt, fill=blue!6, align=left, inner sep=3pt, minimum width=38mm},
  arrow/.style={-stealth, thick, draw=black!60},
  every node/.style={align=left}
]

\node[draw, rounded corners=1pt, fill=gray!5, anchor=north west, inner sep=6pt, font=\ttfamily\footnotesize] (bg) at (0,0) {%
\begin{minipage}{0.97\textwidth}
\colorbox{gray!5}{%
\begin{minipage}{0.99\textwidth}
\centering
\textbf{\normalsize Control (no additional message)}
\end{minipage}%
}\\[2pt]
\colorbox{gray!25}{%
\begin{minipage}{0.99\textwidth}
The election is in 11 days.\\
As someone with your background and values, consider realistically how likely you are to vote in the upcoming election:\\
On a scale from 0 to 4, where:\\
0 - Definitely will not vote\\
1 - Probably will not vote\\
2 - Might or might not vote\\
3 - Probably will vote\\
4 - Definitely will vote\\[2pt]
In your last response (round 13), your voting likelihood was: 2/4.
\end{minipage}%
}\\[5pt]
\colorbox{gray!5}{%
\begin{minipage}{0.99\textwidth}
\centering
\textbf{\normalsize Informational Message}
\end{minipage}%
}\\[2pt]
\colorbox{green!10}{%
\begin{minipage}{0.99\textwidth}
The election is in 12 days. VOTE OR BE SILENCED! One ballot = one voice. Use yours.\\
As someone with your background and values, consider realistically how likely you are to vote in the upcoming election:\\
On a scale from 0 to 4, where:\\
0 - Definitely will not vote\\
1 - Probably will not vote\\
2 - Might or might not vote\\
3 - Probably will vote\\
4 - Definitely will vote\\[2pt]
In your last response (round 13), your voting likelihood was: 2/4.
\end{minipage}%
}\\[5pt]
\colorbox{gray!5}{%
\begin{minipage}{0.99\textwidth}
\centering
\textbf{\normalsize Social Message}
\end{minipage}%
}\\[2pt]
\colorbox{green!20}{%
\begin{minipage}{0.99\textwidth}
The election is in 12 days. VOTE OR BE SILENCED! One ballot = one voice. Use yours. 14625 users have indicated they're likely to vote. From users you follow, users 3677, 12604, 11932, 16785, 949, 14726 are planning to vote. Will you?\\
As someone with your background and values, consider realistically how likely you are to vote in the upcoming election:\\
On a scale from 0 to 4, where:\\
0 - Definitely will not vote\\
1 - Probably will not vote\\
2 - Might or might not vote\\
3 - Probably will vote\\
4 - Definitely will vote\\[2pt]
In your last response (round 13), your voting likelihood was: 2/4.
\end{minipage}%
}
\end{minipage}%
};

\end{tikzpicture}
\caption{Three Experimental scenarios with different banners in the voting intention survey.}
\label{fig:treatment-message}
\end{figure}

\bibliography{mypaper}
\bibliographystyle{apalike}

\end{document}